\documentclass[prb,aps,twocolumn,showpacs,amsmath,amssymb,floatfix,citeautoscript]{revtex4-1}
\usepackage{graphicx}
\usepackage{amsmath}
\usepackage{amssymb}
\usepackage{times}
\usepackage{color}

\newcommand{\Rmnum}[1]{\expandafter\@slowromancap\romannumeral #1@}

\usepackage{color}

\begin{document}

\title{
Tunable Quantum Phase Transitions in a
Resonant Level Coupled to Two Dissipative Baths
}

\author{Dong E. Liu, Huaixiu Zheng, Gleb Finkelstein, and Harold U. Baranger}
\affiliation{
Department of Physics, Duke University, Box 90305, 
Durham, North Carolina 27708-0305, USA}

\begin{abstract}
We study tunneling through a resonant level connected to two dissipative bosonic baths: one is the resistive environment of the source and drain leads, while the second comes from coupling to potential fluctuations on a resistive gate. We show that several quantum phase transitions (QPT) occur in such a model, transitions which emulate those found in interacting systems such as Luttinger liquids or Kondo systems. We first use bo\-son\-ization to map this dissipative resonant level model to a resonant level in a Luttinger liquid, one with, curiously, two interaction parameters. Drawing on methods for analyzing Luttinger liquids at both weak and strong coupling, we obtain the phase diagram. For strong dissipation, a Berezinsky-Kosterlitz-Thouless QPT separates strong-coupling and weak-coupling (charge localized) phases. In the source-drain symmetric case, all relevant backscattering processes disappear at strong coupling, leading to perfect transmission at zero temperature. In fact, a QPT occurs as a function of the coupling asymmetry or energy of the resonant level: the two phases are (i) the system is cut into two disconnected pieces (zero transmission), or (ii) the system is a single connected piece with perfect transmission, except for a disconnected fractional degree of freedom. The latter arises from the competition between the two fermionic leads (source and drain), as in the two-channel Kondo effect. 
\end{abstract}

\date{January 20, 2014}

\maketitle

\section{Introduction}

In a resonant level system, quantum tunneling combined with dissipation gives rise to quantum phase transitions (QPT). The effect of dissipation caused by the environment on quantum tunneling is, of course, a classic topic in the foundations of quantum mechanics \cite{Feynman-Vernon,Zurek:2003}.
In the case of quantum tunneling, the dissipative bosonic modes of the environment generally suppress the tunneling rate, with the degree of suppression depending on the bosonic density of states and the coupling strength \cite{LeggettRMP87}. Experimentally, tunneling with dissipation can be readily realized in a tunnel barrier contacted by resistive leads \cite{IngoldNazarov92,FlensbergRev92}. The electromagnetic excitations in the leads provide a bosonic bath with a linear density of states (Ohmic environment); the coupling strength $r=e^2 R_e/h$ is determined by the lead ({\it i.e.}\ environmental) resistance $R_e$. The key experimental observable is the electrical conductance through the barrier \cite{DelsingPRL89,GeerligsEPL89,ClelandPRL90,KuzminPRL91,PopovicPRB93,PekolaPRL96,ZhengSSCom98,PierrePRL01,Bomze09,Mebrahtu12}, which as a function of temperature $T$ exhibits a power law \emph{suppression} $G\propto  T^{2r}$.
In contrast, in the resonant level system that we study, the conductance is not always suppressed by the environment; the transition between the strong tunneling and suppressed tunneling regimes was shown to be a QPT \cite{Mebrahtu12,Mebrahtu13}. 

Quantum phase transitions have been extensively investigated in a variety of
contexts \cite{SachdevBook,CarrBook,SiIngersent01,VojtaPhilMag06}. In nanoscale systems, it is appropriate to consider \emph{boundary} QPT, which denotes a QPT due to the boundary degrees of freedom (such as, for instance, a spin or single fermionic state) \cite{VojtaPhilMag06}. 
In recent years, there have been three experiments in quantum dot systems that show clear evidence of a QPT \cite{Potok2CK07,Roch08,Mebrahtu12}.
Quantum dots connected to leads are a natural place to look for boundary QPT because of their tunability and flexibility. Indeed, theoretically, many realizations of boundary QPT have been proposed using quantum dots:\cite{VojtaPhilMag06,Hofstetter01,BordaSU4PRL03,OregPRL03,Garst04,PustilnikPRB04,LeHurPRL04,Galpin05,BordaZarandX06,ZarandPRL06,ChungQPT,UlloaPRL09,GoldsteinPRL10,LiuPRL10} 
in multi-dot and multi-level systems, competition between different interactions involving the boundary degree of freedom (dot-lead Kondo interaction, dot-dot or level-level exchange interaction, or Coulomb electrostatic interaction, for instance) produces QPT. Boundary QPT also occur in pseudo-gap Kondo or Anderson models \cite{VojtaPhilMag06}, which could be realized using a quantum dot and a nanoscale Aharonov-Bohm interferometer \cite{UlloaPRL09}. 
Finally, for our purposes it is important to note that boundary QPT can be caused by dissipation: coupling a boundary degree of freedom to an environment causes a qualitative change in behavior for sufficiently strong coupling. Transitions of this type were among the first QPT to be studied in detail \cite{Chakravarty82,LeggettRMP87}, in the form of the ``spin-boson model'' in which there is a transition from a phase in which the spin flips to one in which it is frozen. 

Tunneling with dissipation is closely related to tunneling in a Luttinger liquid (a one-dimensional system with electron-electron interactions \cite{GiamarchiBook}). This appears natural since dissipation connected to the environmental resistance is caused by the electron charge coupling to electromagnetic modes of the environment, thus making a link to the plasmon modes of the Luttinger liquid. For tunneling through a single barrier, a mapping between the two problems makes the connection explicit \cite{SafiSaleur04}. Such a mapping can also be made for our problem of tunneling through a resonant level, as we have shown previously \cite{Mebrahtu12,Mebrahtu13}. This allows us to draw on the extensive literature on resonant tunneling in a Luttinger liquid \cite{Kane92,*Kane92a,EggertAffleck92,Furusaki93,Sassetti_Napoli_Weiss_95,FendleyPRB95,*FendleyPRL95,MatveevPRB95,FurusakiMatveevPRB95,Furusaki98,FurusakiMatveevPRL02,NazarovGlazman03,polyakov03,komnik03,Meden2005,EnssPRB05,goldstein09,GoldsteinPRL10a}, 
in which, in particular, QPT are  known to occur.

\begin{figure}[t]
\centering
\includegraphics[width=3.0in,clip]{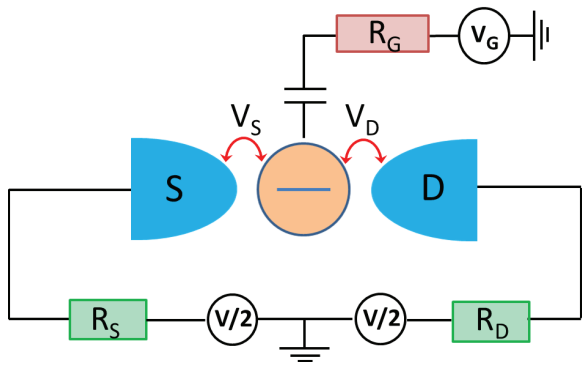}
\caption[a spinless quantum dot is coupled to two resistive leads.]
{Schematic of a spinless quantum dot coupled to two conducting leads and a gate. The source and drain junctions are characterized by tunneling amplitudes $V_{S}$ and $V_{D}$, as well as capacitances $C_S$ and $C_D$. 
The dot-leads system is symmetrically biased by a voltage $V$ through the lead resistances $R_S$ and $R_D$. 
The gate is capacitively coupled to the dot (capcitance $C_G$) through a resistance $R_G$. We consider the simplified situation in which 
$C_S = C_D \equiv C$ and $R_S = R_D \equiv R/2$.}
\label{fig:setup}
\end{figure}

Here we study tunneling through a resonant level which is coupled to two dissipative baths: one produced by the resistive source and drain leads, and a second connected to a gate potential that shifts the energy of the resonant level (see Fig.\,\ref{fig:setup}). While coupling a resonant level to one or the other type of bath has been considered previously 
\cite{IngoldNazarov92,Nazarovbook,IngoldPhysicaB1991,ImamAverinPRB94,ButtikerPRL00,LeHurPRL04,LeHurLiPRB05,BordaPRB05,BordaZarandX06,GlossopIngersent07,Florens07,ChungQPT,ChengIngersent09}, 
this is, as far as we know, the first study in which both types of bath are treated on equal footing. 

Two types of QPT are shown to exist in this system: One involves freezing of the charge fluctuations on the level---it is analogous to the localization transition in the spin-boson model mentioned above \cite{Chakravarty82,LeggettRMP87}---and is well-known to be of the Berezinsky-Kosterlitz-Thouless type. A second transition is associated with a special point: for symmetric coupling and on resonance, one obtains perfect conductance through the level in contrast to the zero conductance state in all other cases.  Our analysis draws on and is analogous to that for tunneling through a resonant level in a Luttinger liquid. However, the mapping presented below shows that the presence of two dissipative baths produces notable differences, differences that we emphasize. These results deepen the close link established in earlier work \cite{Kane92,MatveevGlazman93,Furusaki93,SassettiWeissEPL94,SafiSaleur04,LeHurLiPRB05,Florens07,ParmentierPierre11,Mebrahtu12,JezouinPierre13,Mebrahtu13} 
between effects produced by dissipation and those caused by electron-electron interactions. Indeed, coupling to dissipation can be used to emulate what happens in a strongly interacting electron system \cite{Mebrahtu12,JezouinPierre13,Mebrahtu13}.
Since the type of system we study is very flexible and can be extended, for instance, to several quantum dots connected in a variety of ways to leads and gates, this suggests the possibility of using dissipative systems as a quantum simulator of strongly correlated electronic phenomena. 

The structure of the rest of the paper is as follows. In Sec.\,\ref{sec:model_dissipative}, we introduce a resonant level model that incorporates two types of dissipative baths: one couples to the tunneling process while the other couples to the voltage fluctuations of the dot. Sec.\,\ref{sec:bosonization} shows how the model can be rewritten using bosonization in order to incorporate the environmental contribution into the bosonic fields describing the leads; the corresponding transformations of the current operator are explicitly discussed in Sec.\,\ref{sec:currentop}. In Sec.\,\ref{sec:mapping_LL}, a mapping is established from our dissipative resonant level model to a model with a resonant level coupled to two Luttinger liquid leads. The phase diagram is obtained in Sec.\,\ref{sec:qpt} through a weak-coupling renormalization group analysis in a Coulomb-gas representation combined with a strong-coupling analysis. In Sec.\,\ref{sec:sequential}, we analyze the sequential tunneling regime. Finally, Sec.\,\ref{sec:conclusion} contains a summary and concluding discussion.

\section{Model: A Dissipative Resonant Level}
\label{sec:model_dissipative}

We study a dissipative resonant level model appropriate for describing a spin-polarized quantum dot coupled to two conducting leads in the presence of an ohmic dissipative environment, as shown in Fig.\,\ref{fig:setup}. Charge fluctuations associated with the dot are coupled to the electromagnetic environment modeled by the three resistors; note that we include dissipation coming from both the gate and the transport leads. At sufficiently low temperature, these charge fluctuations must be treated quantum mechanically. The barriers from the dot to the source and drain are characterized by capacitances as well as tunneling amplitudes. For simplicity we take the capacitance of the source barrier to be the same as that of the drain, both denoted $C$; the resistances connected to the source and drain are likewise equal with value $R/2$ each. (This case is appropriate to describe the experiments in Refs.\,\onlinecite{Mebrahtu12,Mebrahtu13}.) The capacitance and resistance associated with the gate, $C_G$ and $R_G$, can be different.  

The Hamiltonian can be divided into four terms,
\begin{equation}
H=H_{\textrm{Dot}}+H_{\textrm{Leads}}+H_{\textrm{T}}+
H_{\textrm{Env}} \;,
\label{eq:H}
\end{equation}
corresponding, respectively, to the dot, the leads, the tunneling between them, and the environmental modes.
The terms to describe the dot and the leads are straightforward: We keep a single state in the dot (electron creation operator $d^{\dagger}$) whose energy level $\epsilon_d$ is shifted by the average voltage on the gate,
\begin{equation}
   H_{\textrm{Dot}}=\epsilon_d d^{\dagger}d  \;.
\label{eq:Hdot_v0}
\end{equation}
The source (S) and drain (D) leads consist of non-interacting electrons described by
\begin{equation}
   H_{\textrm{Leads}}=
   \sum_{\alpha=S,D}\sum_{k}\epsilon_{k}c_{\alpha k}^{\dagger}c_{\alpha k} \;.
\label{eq:Hleads}
\end{equation}

$H_{\textrm{T}}$ describes the tunneling between the dot and the leads; since electrons are charged, this involves not only conversion of a \textit{d} electron into a quasi-particle but also transfer of a charge. The quantum electrical properties of each capacitor connected to the quantum dot are treated by introducing an operator for the charge fluctuations on each capacitor, denoted $Q_S$, $Q_D$, and $Q_G$, as well as their conjugate phase variables $\varphi_S$, $\varphi_D$, and $\varphi_G$, respectively \cite{IngoldNazarov92,Nazarovbook,IngoldPhysicaB1991}. The latter correspond physically to the time-integrated voltage fluctuations across the capacitor. These quantities obey the commutation relations
\begin{equation}
[\varphi_\alpha , Q_{\alpha'}] = ie\,\delta_{\alpha,\alpha'} \quad\quad {\rm for}\; \alpha,\alpha' =S,\;D,\;G \;.
\end{equation}
The tunneling part of the Hamiltonian is, then, 
\begin{eqnarray}
H_{\textrm{T}} & = & V_{S}\sum_{k}(c_{Sk}^{\dagger}e^{-i\varphi_S}d + {\rm H.c.}) \nonumber \\
& + & V_{D}\sum_{k}(c_{Dk}^{\dagger}e^{-i\varphi_D}d + {\rm H.c.}),
\label{eq:HT_v1}
\end{eqnarray}
where $V_S$ and $V_D$ are the tunnel couplings to, respectively, the source and drain leads. In describing the effect of the dissipative environment by using a single phase factor per junction in the tunneling Hamiltonian, we are neglecting transitions between different momentum states within the same lead, and thus neglecting electron relaxation and decoherence \cite{Nazarovbook}. This approach appears to be adequate if the electromagnetic field propagates much faster than the electrons \cite{Nazarovbook}, which is the case for the samples we have in mind \cite{Mebrahtu12,Mebrahtu13}. A similar model has been used, for instance, in previous work on a resonant level \cite{ImamAverinPRB94}, for a quantum dot in the Kondo regime connected to resistive leads \cite{Florens07}, and for a dissipative dot coupled to a Luttinger liquid \cite{LeHurLiPRB05}. 

To incorporate the effects of the environment, it is convenient, first, to rotate  the charge and phase variables to the following set:
\begin{eqnarray}
Q_1 & = & Q_S + Q_D + Q_G
\\
\varphi_1 & = & 
   \big(\varphi_S + \varphi_D +\frac{C_G}{C}\varphi_G \big)\frac{C}{C_\Sigma}  \\
Q_2 & = & \frac{1}{2} (Q_S - Q_D) \\ 
\varphi_2 & = & \varphi_S - \varphi_D \\
Q_3 & = & \big(\frac{Q_S}{2} + \frac{Q_D}{2} - \frac{C}{C_G}Q_G \big)\frac{C_G}{C_\Sigma} \\
\varphi_3 & = & \varphi_S + \varphi_D -2\varphi_G \;,
\end{eqnarray}
where $C_\Sigma \equiv 2C + C_G$ is the total capacitance of the dot. 
The rotation preserves the canonical commutation relations
\begin{equation}
[\varphi_i , Q_{i'}] = ie\,\delta_{i,i'} \quad\quad {\rm for}\; i,i' =1,\;2,\;3 \;.
\end{equation}
These variables have a natural physical interpretation. First, $Q_1$ is clearly the total charge on the dot, and therefore the operator $e^{i\varphi_1}$ changes this total charge by $e$ \cite{IngoldNazarov92}. Second, $e^{i\varphi_2}$ moves a charge from the source capacitor to the drain capacitor. It thus moves charge around the lower loop in our circuit Fig.\,\ref{fig:setup}. The remaining variable must be orthogonal to the first two. It corresponds to moving charge $2e$ from the source and drain capacitors to the gate, that is, moving charge vertically in our circuit Fig.\,\ref{fig:setup}.

In terms of these rotated coordinates, the tunneling Hamiltonian takes the form
\begin{eqnarray}
H_{\textrm{T}} & = & V_{S}\sum_{k}c_{Sk}^{\dagger}
\exp\big[-i\big(\varphi_1 + \frac{1}{2}\varphi_2 + \frac{1}{2}\frac{C_G}{C_\Sigma}\varphi_3 \big) \big] d  
\nonumber \\
& + & V_{D}\sum_{k} c_{Dk}^{\dagger}
\exp\big[-i\big(\varphi_1 - \frac{1}{2}\varphi_2 + \frac{1}{2}\frac{C_G}{C_\Sigma}\varphi_3 \big) \big]d 
\nonumber \\
& + & {\rm H.c.}
\label{eq:HT_v2}
\end{eqnarray}

It is the coupling of the charge fluctuations to the ohmic environment represented by the resistors which leads to dissipation. A phase variable connected to charge flow through a certain resistance is coupled to the environment represented by that resistance. Thus variable $\varphi_2$ is coupled to an environment characterized by the resistance 
\begin{equation} \label{eq:R2def}
R_2 \equiv R_S +R_D = R \;,
\end{equation}
and $\varphi_3$ is coupled to an environment with dissipation given by 
\begin{equation}\label{eq:R3def}
R_3 \equiv R + 4R_G \;.
\end{equation}
Note that the fact that $\varphi_3$ moves two charges through the gate circuit causes a factor of 4 in the corresponding resistance---dissipation is proportional to the square of the current. Finally, notice that the total charge mode, $(Q_1,\varphi_1)$, does not couple to the environment~\cite{IngoldNazarov92,Nazarovbook}. The reason for this lack of coupling is that the charge involved in fluctuations of $Q_1$ is balanced among the three capacitors: they do not require flow in the external circuit and so do not cause dissipation.


The ratio of the resistance to the quantum of resistance, $R_Q \!=\! h/e^2$, is the key physical quantity, as we will see below. For the various resistances here, this ratio is denoted by $r\!\equiv\!R/R_Q$, $r_G\!\equiv\!R_G/R_Q$, $r_2\!\equiv\!R_2/R_Q$, and $r_3\!\equiv\!R_3/R_Q$. 

Because the two fluctuating modes $(Q_2,\varphi_2)$ and $(Q_3,\varphi_3)$ are orthogonal, we can take their environments to be independent. Each of the phase operators $\varphi_2$ and $\varphi_3$ is coupled to the environment in the usual way:\cite{Devoret97} The resistance is modeled by an infinite collection of LC oscillators which act as a bath; the impedance of the bath viewed from the quantum dot is chosen to match the resistance in the circuit. The phase of each oscillator is bilinearly coupled to the appropriate $\varphi_i$. Upon integrating out the harmonic bath degrees of freedom, the key property is that the decay of the correlation of $\varphi_i$ at long times is \cite{IngoldNazarov92}
\begin{equation}
 \left\langle e^{i\varphi_i(t)}e^{-i\varphi_i(0)}\right\rangle \to 
  \frac{A}{(\omega_{R_i}\,t)^{2r_i} },
 \quad {\rm with}\;i=2 \;{\rm or}\; 3 
 \label{eq:corvarphi}
\end{equation}
where $\omega_{R_i}\!=\!1/(R_{i} C)$ serves as a high energy cutoff and $A$ is a constant.
In this way, one arrives at the natural result that the resistance associated with a given charge fluctuation mode controls its relaxation. In the absence of an environment, $r_i\!=\!0$, the fluctuations are not damped. 

While previously the effect on resonant tunneling of either transport charge fluctuations or gate charge fluctuations have been independently studied \cite{IngoldNazarov92,Nazarovbook,IngoldPhysicaB1991,ImamAverinPRB94,ButtikerPRL00,LeHurPRL04,LeHurLiPRB05,BordaPRB05,BordaZarandX06,GlossopIngersent07,Florens07,ChungQPT,ChengIngersent09},
this is, as far as we know, the first treatment where both charge fluctuation modes have been included on the same footing. As both modes are, of course, present in experiment \cite{Mebrahtu12,Mebrahtu13}, their mutual effects may be important for determining the phases and behavior of the system.

\section{Combining Environment and Leads}\label{sec:bosonization}

In this section, we treat the two leads using bosonic fields so that they may be combined with the phase factors describing the coupling to the environment, following closely the previous literature for tunneling through a single barrier \cite{SafiSaleur04} or quantum dots \cite{LeHurLiPRB05,Florens07}.
Because the dot couples to each lead at a single point, the two metallic leads may be reduced to two semi-infinite one-dimensional free fermionic baths \cite{KaulPRB03,ChoPRB03,SimonAffleckPRB03}. By unfolding the two semi-infinite fermionic fields, one obtains two chiral free fermionic fields; for each of these, we take the point of coupling to the dot to be $x=0$. These chiral fields can be bosonized 
\cite{senechal,GiamarchiBook}, yielding 
\begin{equation}
 c_{S,D}(x)=\frac{1}{\sqrt{2\pi a}}F_{S,D}\exp[i\phi_{S,D}(x)] \;.
\end{equation}
Here, $\phi_{S,D}$ are the bosonic fields, $F_{S,D}$ are the Klein factors needed to preserve the fermionic anticommutation relations, and $a$ is the short time cutoff. 
The commutation relations for these chiral bosonic fields are
\begin{equation}
 [\partial_x \phi_{i}^{0}(x),\,\phi_{j}^{0}(x')] = i\delta_{ij}\pi\, \delta (x-x'),\quad i,j=S,D .
\end{equation}

We now rotate the lead basis by introducing the flavor field $\phi_{f}^{0}$ and charge field $\phi_{c}^{0}$, 
\begin{equation}
\phi_{f}^{0} = \frac{\phi_{S}-\phi_{D}}{\sqrt{2}} \quad\textrm{and}\quad
\phi_{c}^{0} = \frac{\phi_{S}+\phi_{D}}{\sqrt{2}},
\label{eq:CF_field}
\end{equation}
in terms of which the lead Hamiltonian is simply
$H_{\textrm{Leads}}=\frac{v_{F}}{4\pi}\int_{-\infty}^{\infty}dx \big[\left(\partial_{x}\phi_{c}^{0}\right)^{2}+\big(\partial_{x}\phi_{f}^{0}\big)^{2}\big]$ since it is non-interacting. 
The tunneling Hamiltonian Eq.\,(\ref{eq:HT_v2}) becomes
\begin{eqnarray}
H_{\textrm{T}}  & =&   V_{S}\frac{F_{S}}{\sqrt{2\pi a}}\,
e^{-\frac{i}{\sqrt{2}}[\phi^0_c+\phi^0_f](x=0)}
e^{-i(\varphi_1 + \frac{1}{2}\varphi_2 + \frac{1}{2}\frac{C_G}{C_\Sigma}\varphi_3)}d 
\nonumber \\
  & + & V_{D}\frac{F_{D}}{\sqrt{2\pi a}}\,e^{-\frac{i}{\sqrt{2}}[\phi^0_c-\phi^0_f](x=0)}
e^{-i(\varphi_1 - \frac{1}{2}\varphi_2 + \frac{1}{2}\frac{C_G}{C_\Sigma}\varphi_3)}d
\nonumber \\
& + & {\rm H.c.} . 
\label{eq:HT_Bosonization}
\end{eqnarray}
Note that $\phi_{f}^{0}(x\!=\!0)$ and $\phi_{c}^{0}(x\!=\!0)$ enter in a way very similar to that of $\varphi_2$ and $\varphi_3$. Indeed, since both the correlation functions of $\varphi_2$ and $\varphi_3$ [Eq.\,(\ref{eq:corvarphi})] and those of the free chiral fields \cite{GiamarchiBook} describing the leads have a power law decay in time, we shall be able to combine $\varphi_2$ with the flavor field $\phi_{f}^{0}(x \!=\! 0)$ and likewise combine $\varphi_3$ with the charge field $\phi_{c}^{0}(x \!=\! 0)$. At this point we drop $\varphi_1$ from our expressions since it is not coupled to the environment and so plays no role. 

To combine the phase factors in the desired way, an analytic continuation is needed: the environment phase factor $\varphi_2$ is defined only on the time axis whereas the field $\phi_{f}^{0}$ depends on both space and time. We take $\varphi_2(t) \rightarrow \varphi_2(t,x)$ and extend the correlation function to the full space with the commutation relation
\begin{equation}
 [\partial_x\varphi_{i}(x),\varphi_{j}(x')]=i \,2r_{i}\delta_{ij}\pi \delta(x-x'),\quad i,j = 2,3.
\end{equation}
Note that this continuation dose not influence the physics because the tunneling involves the phase only at $x\!=\!0$. Now, $\varphi_2$ can be absorbed by $\phi_{f}^{0}$ by redefining the fields as
\begin{eqnarray}
\phi_{f} & \equiv & \sqrt{g_{f}}\Big(\phi_{f}^{0}+\frac{1}{\sqrt{2}}\varphi_2\Big),\nonumber \\
\varphi_f' & \equiv & \sqrt{g_{f}}\Big(\sqrt{r}\phi_{f}^{0}
-\frac{1}{\sqrt{2r}}\varphi_2 \Big),
\label{eq:absorption_f}
\end{eqnarray}
where 
\begin{equation}
g_{f} \equiv \frac{1}{1+r} \leq 1  \;.  \label{eq:gfdef}
\end{equation} 
In a similar way, the phase operator $\varphi_3$ can be absorbed by
the charge field $\phi_{c}^{0}$ through the transformation
\begin{eqnarray}
\phi_{c} & \equiv & \sqrt{g_{c}}\Big(\phi_{c}^{0}+\frac{1}{\sqrt{2}} 
\frac{C_G}{C_\Sigma} \varphi_3\Big),\nonumber \\
\varphi_c' & \equiv & \sqrt{g_{c}}\Big(\frac{C_G}{C_\Sigma}\sqrt{r_3}\phi_{c}^{0}-\frac{1}{\sqrt{2r_3}} \varphi_3\Big),
\label{eq:absorption_c}
\end{eqnarray}   
where 
\begin{equation}
g_{c} \equiv \frac{1}{1+\big(\frac{C_G}{C_\Sigma}\big)^2 r_3} \leq 1 \;.
 \label{eq:gcdef}
\end{equation} 
The prefactors in these transformations are uniquely determined by the requirement that the new fields obey canonical commutation relations:
\begin{align}
 [\partial_x \phi_{i}(x),\,\phi_{j}(x')] &= i\pi\, \delta_{ij}\,\delta (x-x'),\quad i,j=c,f \;,\nonumber\\
 [\partial_x \varphi_i'(x),\,\varphi_j'(x')] &= i\pi\, \delta_{ij}\,\delta (x-x'), \nonumber\\
 [\varphi_i'(x),\,\phi_{j}(x')] &= 0  \;.
\end{align}
In terms of these fields, the Hamiltonian becomes
\begin{eqnarray}
 H&=&  H_{\textrm{Dot}} + \frac{v_{F}}{4\pi}\int_{-\infty}^{\infty}dx\left[\left(\partial_{x}\phi_{c}\right)^{2}+\left(\partial_{x}\phi_{f}\right)^{2}\right] + H_{\textrm{Env}}
 \nonumber\\
 & + & V_S \Big[\frac{F_{S}}{\sqrt{2\pi a}} e^{-i\frac{\phi_{f}(x=0)}{\sqrt{2 g_f}}}  e^{-i\frac{\phi_{c}(x=0)}{\sqrt{2 g_{c}}}}  d + {\rm H.c.}\Big] \nonumber\\
 & + & V_D \Big[\frac{F_{D}}{\sqrt{2\pi a}}  e^{i\frac{\phi_{f}(x=0)}{\sqrt{2 g_{f}}}}  e^{-i\frac{\phi_{c}(x=0)}{\sqrt{2 g_{c}}}}  d + {\rm H.c.}\Big] \;;
\label{eq:heff_tunneling}
\end{eqnarray}
the new phase fluctuations $\varphi_f'$ and $\varphi_c'$ decouple from the dot and tunneling term, and so we omit them.

Because of the coefficients in the exponentials for the tunneling Hamiltonian, \emph{the transformed fields are effectively interacting}: the dissipative environment (the phase factors $\varphi_2$ and $\varphi_3$) is incorporated in the new flavor and charge fields $\phi_{f}$ and $\phi_{c}$ at the expense of introducing interaction parameters $g_{f}$ and $g_{c}$. A similar mapping was obtained for a quantum dot in the Kondo regime in Ref.\,\onlinecite{Florens07} and for a dissipative dot coupled to a single chiral Luttinger liquid in Ref.\,\onlinecite{LeHurLiPRB05}. 
The Hamiltonian Eq.\,(\ref{eq:heff_tunneling}) is indeed a Luttinger liquid model, but a somewhat unusual one in which the dot couples to two Luttinger liquids with different interaction parameters. Notice that in the limit $C_G \ll C$ relevant for the experiment of Refs.\,\onlinecite{Mebrahtu12,Mebrahtu13}, one has $g_c=1$. In presenting below the properties implied by this Hamiltonian, we shall in particular emphasize features connected to the fact that the two interaction parameters are different from each other.

\section{Current Operator}\label{sec:currentop}

The representation in Eq.\,(\ref{eq:heff_tunneling}) is convenient for obtaining the partition function and so thermodynamic quantities (see Section\,\ref{sec:qpt});  however, transport properties, such as the current through the resonant level, may be affected by unitary transformations. We therefore check how the current operator transforms in the operations used to arrive at Eq.\,(\ref{eq:heff_tunneling}).

In the first step, two metallic leads were reduced to two chiral free fermionic fields $c_{S,D}(x)$, with the resonant level coupling to $c_{S,D}(0)$. Due to the linear dispersion of the chiral fermions, the current operator can be written as the difference between the densities of the incoming and outgoing electrons in either the $S$ or $D$ channel \cite{komnik03,Egger&Grabert98}:
\begin{equation}
 I_{S,D}=e v_{F} \big[ c_{S,D}^{\dagger}c_{S,D} (x\rightarrow \infty) - c_{S,D}^{\dagger}c_{S,D} (x\rightarrow -\infty)  \big].
\end{equation}
One can rewrite the density operators in terms of the bosonic fields, $c_{S,D}^{\dagger}(x)c_{S,D}(x)=\partial_x \phi_{S,D}(x)/2\pi$,
yielding
\begin{equation}
 I_{S,D}=\frac{e v_{F}}{2\pi} \big[ \partial_x \phi_{S,D}(\infty) - \partial_x \phi_{S,D}(-\infty)  \big]\;.
 \label{eq:currentSD}
\end{equation}
Since the current obeys $I=\alpha I_S -(1-\alpha) I_D$ for any $0\leq \alpha\leq 1$, the current operator in the $\phi_{c,f}^0$ basis [Eq.\,(\ref{eq:CF_field}] is
\begin{equation}
 I=\frac{e v_F}{2\sqrt{2}\pi} \big[ \partial_x \phi_{f}^0(\infty) - \partial_x \phi_{f}^{0}(-\infty)  \big]\;.
\end{equation}
The charge field does not contribute to the current. 

The current operator is potentially affected by the transformation Eq.\,(\ref{eq:absorption_f}) used to absorb the the environment phase factor $\varphi_2$. The current operator in the new basis is 
\begin{eqnarray}
 I &=&  \frac{e v_F}{2\sqrt{2}\pi} \sqrt{g_f}  \big[ \partial_x \phi_{f}(\infty) - \partial_x \phi_{f}(-\infty)  \big] \nonumber\\
   &  +& \frac{e v_F}{2\sqrt{2}\pi} \sqrt{g_f}\sqrt{r}   \big[ \partial_x \varphi_f'(\infty) - \partial_x \varphi_f'(-\infty)  \big]
\end{eqnarray}
Since the phase fluctuation field $\varphi_f'$ decouples from the other parts of the system, its contribution to the current vanishes: $\partial_x \varphi_f'(\infty) - \partial_x \varphi_f'(- \infty)=0$. Thus, the current operator in the final transformed basis depends only on the $\phi_f$ field,
\begin{equation}
 I=\frac{e v_F}{2\sqrt{2}\pi} \sqrt{g_f}  \big[ \partial_x \phi_{f}(\infty) - \partial_x \phi_{f}(-\infty)  \big];
\label{eq:current}
\end{equation}
we recognize the current operator \cite{komnik03,Egger&Grabert98} for a chiral Luttinger liquid (up to a factor of $\sqrt{2}$).


\section{Mapping to Physical Luttinger Liquid Model}\label{sec:mapping_LL}

The Hamiltonian Eq.\,(\ref{eq:heff_tunneling}) does not, unfortunately, directly describe an electron hopping between the quantum dot and real physical leads, in particular because of the presence of a three body interaction term in $H_{\textrm{T}}$. Thus it is interesting to develop an alternative physical model.

To obtain a physical model, we wish to eliminate the three-body interaction in the Hamiltonian Eq.\,(\ref{eq:heff_tunneling}). In order to combine the two fields $\phi_c$ and $\phi_f$ in the exponents of the tunneling term, their coefficients must be the same. We can change the coefficient of the $\phi_c$ term so that this is true by applying the unitary transformation \cite{komnik03,emery92}
\begin{equation}
U=\exp \Big[ i \Big(\frac{1}{\sqrt{2g_{c}}} - \frac{1}{\sqrt{2g_{f}}} \Big)
\big(d^{\dagger}d-1/2\big) \phi_c(0) \Big] ,
\end{equation}
at the cost of introducing a density-density interaction term between the leads and the dot. 
As for any unitary transformation of the form $\exp[i \alpha (d^{\dagger}d-1/2) \phi_c(0)]$, $U$ commutes with the current operator \cite{komnik03,emery92} and so does not affect the current. After applying this transformation and redefining new ``source'' and ``drain'' channels by 
\begin{equation}
 \widetilde{\phi}_{S} = \frac{\phi_{c}+\phi_{f}}{\sqrt{2}} \quad\textrm{and}\quad
 \widetilde{\phi}_{D} = \frac{\phi_{c}-\phi_{f}}{\sqrt{2}},
\end{equation}
the Hamiltonian becomes
\begin{widetext}
\begin{eqnarray}
 \widetilde{H}  =  U^{\dagger} H U 
  & = & H_0 + V_S \Big[\frac{F_{S}}{\sqrt{2\pi a}}e^{-i\frac{\widetilde{\phi}_{S}(x=0)}{\sqrt{g_{f}}}} d + {\rm H.c.}\Big]
 + V_D \Big[\frac{F_{D}}{\sqrt{2\pi a}}e^{-i\frac{\widetilde{\phi}_{D}(x=0)}{\sqrt{g_{f}}}}d + {\rm H.c.} \Big]\nonumber\\
 & & + \frac{v_{F}}{4}\Big(\frac{1}{\sqrt{g_c}}-\frac{1}{\sqrt{g_f}}\Big) (d^{\dagger}d-1/2) 
  \Big[\partial_x \widetilde{\phi}_{S}(x=0)+\partial_x \widetilde{\phi}_{D}(x=0) \Big], 
\label{eq:SC_H}
\end{eqnarray}
\end{widetext}
where the last term is the density-density interaction produced by the unitary transformation. 
One finds that the current is given by 
the usual expression for an interacting chiral field \cite{komnik03,Egger&Grabert98}, 
$ I =\frac{e v_{F}}{2\pi} \sqrt{g_f} [ \partial_x \widetilde{\phi}_{S,D}(\infty) - 
\partial_x \widetilde{\phi}_{S,D}(-\infty) ] $,
in terms of these effective source and drain channels. 

Since a level coupling to a chiral Luttinger liquid is equivalent to a level coupling to the end of a non-chiral luttinger liquid \cite{GiamarchiBook}, the original model is thus mapped to a very natural physical system: a resonant level embedded in a Luttinger liquid having a \emph{single} interaction parameter $g_f=1/(1+r)$ with an additional electrostatic interaction between the dot and the ends of the two leads [last term in (\ref{eq:SC_H})]. 
If the values of the resistances and capacitances are carefully chosen so that $g_c=g_f$, this extra electrostatic interaction vanishes. The model is then exactly equivalent to a double barrier in a spinless Luttinger liquid, a situation which has been intensively studied \cite{Kane92,*Kane92a,EggertAffleck92,Furusaki93,Sassetti_Napoli_Weiss_95,FendleyPRB95,*FendleyPRL95,MatveevPRB95,FurusakiMatveevPRB95,Furusaki98,FurusakiMatveevPRL02,NazarovGlazman03,polyakov03,komnik03,Meden2005,EnssPRB05,goldstein09,GoldsteinPRL10a}.

Another useful representation is obtained by applying a slightly different unitary transformation \cite{emery92,komnik03}, $U'\!=\!\exp[i(d^{\dagger}d-1/2) \phi_c(0) / \sqrt{2g_c}]$, to eliminate the $\phi_c$ field from the tunneling process entirely. As in the previous transformation, an electrostatic density-density interaction between the leads and the dot is generated,
\begin{equation}
 H_{\rm int}=\frac{v_{F}}{2} \Big( 
 -\frac{1}{\sqrt{2g_c}}\Big) (d^{\dagger}d-1/2) \partial_x \phi_c(x=0).
\end{equation}
From this representation, the relation with the two-channel Kondo model, which shows exotic non-Fermi-liquid behavior \cite{LudwigTCK91,AffleckTCK92,AffleckTCK93}, can be made clear \cite{emery92,komnik03}, a situation we studied recently \cite{Mebrahtu13}. 
For $g_f\!=\!1/2$ (i.e., $r\!=\!1$), a refermionization procedure is possible,
$\psi_{f}\!=\!e^{i\phi_{f}}/\sqrt{2\pi a}$. If in addition the density-density interaction term is discarded (even though typically large), one arrives at a non-interacting Majorana resonant level model, which is exactly the same as that reached by using a bosonization procedure \cite{emery92,GogolinBook} in the two-channel Kondo model. The connection between resonant tunneling in a Luttinger liquid and the two-channel Kondo has been extensively investigated \cite{komnik03,GoldsteinPRB10,MatveevPRB95,FurusakiMatveevPRB95,EggertAffleck92}. In contrast, the connection in the context of the dissipative resonant tunneling problem has received limited attention. In Ref.\,\onlinecite{Mebrahtu13} the connection was made explicit and, furthermore, studied experimentally.

\section{Scaling and Quantum Phase Transitions} 
\label{sec:qpt}

Having transformed our problem to a Luttinger liquid form, we can now bring to bear the many techniques developed for problems involving impurities in a Luttinger liquid \cite{GiamarchiBook,SafiSaleur04,Kane92,*Kane92a,EggertAffleck92,Furusaki93,Sassetti_Napoli_Weiss_95,FendleyPRB95,*FendleyPRL95,MatveevPRB95,FurusakiMatveevPRB95,Furusaki98,FurusakiMatveevPRL02,NazarovGlazman03,polyakov03,komnik03,Meden2005,EnssPRB05,goldstein09,GoldsteinPRL10a}. We proceed from the version of our model in Sec.\,\ref{sec:bosonization}, Eq.\,(\ref{eq:heff_tunneling}). First, we develop a ``Coulomb-gas'' representation, then use it to generate a weak-coupling renormalization group (RG) treatment, and finally turn to characterizing the strong-coupling fixed point. Since much of the technical development is well known, we only sketch it briefly here; rather, we concentrate on the results and the differences induced by $g_c \neq g_f$.   

\subsection{Coulomb Gas Representation}

The ``Coulomb-gas'' representation is a convenient way to derive RG equations \cite{AndYuvalHamannPRB70} and has been used for similar problems in, e.g., Refs.\,\onlinecite{Kane92} and \onlinecite{Goldstein10a}. 
We start by expanding the corresponding partition function in powers of the tunneling, $V_{S}$ and $V_{D}$. Since the tunneling acts only at $x=0$, it is convenient to perform a partial trace in the partition function and integrate out fluctuations in $\phi_{c,f}(x)$ for all $x\neq0$ \cite{Kane92,Furusaki93,AndYuvalHamannPRB70}. If in addition one integrates out the environmental modes (they are harmonic), the effective action absent the tunneling is 
\begin{eqnarray}
S_0^{\textrm{eff}} & = & \frac{1}{\beta}\sum_{n}|\omega_{n}|\left(|\phi_{c}(\omega_{n})|^{2}+|\phi_{f}(\omega_{n})|^{2}
\right) \nonumber \\
& & + \int_{0}^{\beta} d \tau \bar d (\partial_{\tau}-\epsilon_d) d,
\label{eq:Seff}
\end{eqnarray}
where $\omega_{n}\!=\!2\pi n/\beta$ are the Matsubara frequencies and the bosonic fields all refer to their $x\!=\!0$ value.
The Lagrangian for the tunneling term follows directly from Eq.\,(\ref{eq:heff_tunneling}),
\begin{eqnarray}
L_{\textrm{T}} & = & -V_{S}\left(\frac{F_{S}}{\sqrt{2\pi a}}e^{-i\frac{1}{\sqrt{2g_{c}}}\phi_{c}(\tau)}e^{-i\frac{1}{\sqrt{2g_{f}}}\phi_{f}(\tau)}d+ {\rm c.c.} \right) \nonumber \\
& & -V_{D}\left(\frac{F_{D}}{\sqrt{2\pi a}}e^{-i\frac{1}{\sqrt{2g_{c}}}\phi_{c}(\tau)}e^{i\frac{1}{\sqrt{2g_{f}}}\phi_{f}(\tau)}d+ {\rm c.c.} \right), 
\quad\quad
\label{eq:LT}
\end{eqnarray}
in terms of which the tunneling action is 
$S_{\textrm{T}} =  \int_{0}^{\beta} L_{\textrm{T}}(\tau) d\tau$.

One expands the partition function,
$Z  =  \int [D\varphi_c] [D\varphi_f] [D d] e^{-S_{0}^{\textrm{eff}}} e^{-S_{\rm T}}$, in terms of $S_{\rm T}$ and evaluates the resulting correlators using $S_0^{\textrm{eff}}$.
The result is a classical one-dimensional (1D) statistical mechanics problem with the
partition function 
\begin{widetext}
\begin{eqnarray}
Z & = & \!\sum_{\sigma=\pm}\sum_{n}\sum_{\{q_{i}=\pm\}}\!V_{S}^{\sum_{i}(1+q_{i}p_{i})/2}V_{D}^{\sum_{i}(1-q_{i}p_{i})/2}
\!\!\intop_{0}^{\beta}\!\!d\tau_{2n}\!\!\intop_{0}^{\tau_{2n}}\!\!d\tau_{2n-1}\ldots 
\!\!\intop_{0}^{\tau_{2}}\!\!d\tau_{1}\exp\{\sum_{i<j}V_{ij}\}\,\exp\{\epsilon_{d}[\beta\frac{1-\sigma}{2}+\sigma\!\!\!\sum_{1\leq i\leq2n}\!p_{i}\tau_{i}]\},\;\;\;\; \\[0.05in]
& & V_{ij} = \frac{1}{2g_{f}}\big[q_{i}q_{j}+K_{1}p_{i}p_{j}+K_{2}(p_{i}q_{j}+p_{j}q_{i})\big]\ln(\frac{\tau_{i}-\tau_{j}}{\tau_{c}})\;. 
\label{eq:ZCgas}
\end{eqnarray}
\end{widetext}
Here, $\tau_c$ is a short-time cutoff, $q_i$ and $p_i$ are two types of charges that take values $\pm1$, 
and $K_1$ and $K_2$ characterize the strength of the logarithmic interactions between the various pairs of charges. Physically, the $q_{i}$ charge represents the way tunneling events contribute to the transport current:  $+1$ denotes an event from source to dot or from dot to drain, while $-1$ is for the reverse processes. The $q_{i}q_{j}$ terms are obtained from correlators of $\phi_f$, which therefore produce $q_{i}q_{j}/2g_f$. On the other hand, the $p_{i}$ charge represents the way tunneling events contribute to the total charge on the dot: $+1$ for tunneling onto the dot from either lead, and $-1$ for tunneling off. The $p_{i}p_{j}$ terms are obtained from correlators of $\phi_c$, which give $p_{i}p_{j}/2g_c$. As there are no cross correlations between $\phi_f$ and $\phi_c$, the $q$ and $p$ charges do not interact initially. Thus the initial, or ``bare'', values of $K_1$ and $K_2$ are
\begin{equation}
 K_{1}^{\text{bare}}=\frac{g_{f}}{g_{c}}=\frac{1+\big(\frac{C_G}{C_\Sigma}\big)^2 r_3}{1+r}
 \;,\qquad
 K_{2}^{\text{bare}}= 0 \;.
\label{eq:K1bare}
\end{equation}
Note that the initial value for $K_1$ here differs from that for resonant tunneling in a Luttinger liquid \cite{Kane92} for which $K_{1}^{\rm bare}\!=\!1$.

A number of constraints should be respected in constructing the charge configurations appearing in the partition function  \cite{AndYuvalHamannPRB70,Kane92,GiamarchiBook}. First, the total system is charge neutral, $\sum_{i}q_{i}=\sum_{i}p_{i}=0$. 
Second, the sign of the $p_{i}$ charge must alternate in time since the dot has only two states, empty or full. This leads to a renormalization of the interaction, $K_1$, between the $p_i$ charges. Finally, for the $q_i$ charge, there is no ordering restriction, and so the interaction between the $q_{i}$ charges, $1/g_f$, does not get renormalized.

The Coulomb gas model that emerges here is the same as that for resonant tunneling in a Luttinger liquid \cite{Kane92}, except that the initial value for the interaction between $p_{i}$ charges is tunable here by changing the dissipative resistances $r$ or $r_3$. In the limit $C_G\ll C$ in which  dissipation from the gate is not present, $K_{1}^{\text{bare}}= 1/(1+r)$. In the opposite limit $C_G\gg C$ in which gate dissipation dominates, $K_{1}^{\text{bare}}=1+ 4r_G/(1+r)$.

The Coulomb gas representation provides a convenient route to the weak-coupling RG equations \cite{GiamarchiBook,Kane92,AndYuvalHamannPRB70}, by integrating out the degrees of freedom between 
$\tau_c$ and $\tau_c+d\tau$. We consider the on-resonance case, $\epsilon_{d}=0$, so that the last term in Eq.\,(\ref{eq:ZCgas}) is equal to 1. 
The resulting RG equations are the same as for resonant tunneling in a Luttinger liquid \cite{Kane92},
\begin{eqnarray}
 \frac{dK_1}{d\ln\tau_c} & = & -4\tau_c^2 [K_1 (V_S^2+V_D^2) + K_2 (V_S^2-V_D^2)] \nonumber \\
 \frac{dK_2}{d\ln\tau_c} & = & -2\tau_c^2 [K_2 (V_S^2+V_D^2) + (V_S^2-V_D^2)] \nonumber \\
 \frac{dV_S}{d\ln\tau_c} & = & V_S [1-\frac{1+r}{4} (1+K_1+2 K_2)] \nonumber \\
 \frac{dV_D}{d\ln\tau_c} & = & V_D [1-\frac{1+r}{4} (1+K_1-2 K_2)] \;.\label{eq:RG_VD}
\end{eqnarray}
Because of the correspondence with resonant tunneling in a Luttinger liquid, we can immediately deduce a great deal about the properties of this system. 

\subsection{Symmetric Barriers and On Resonance: 
A Special Point}
\label{sec:SymmOnres}

\begin{figure}[t]
\begin{centering}
\includegraphics[width=2.8in,clip]{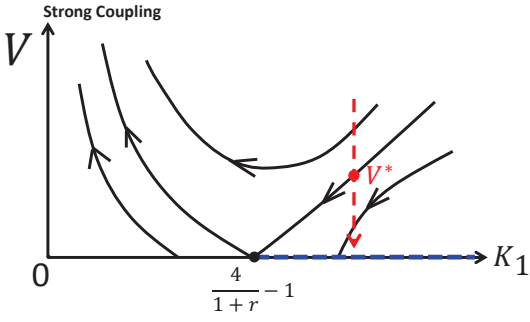}
\par\end{centering}
\caption[Schematic representation of the renormalization group flow for symmetric case.]
{Schematic representation of the RG flow for the symmetric case ($V_S\!=\!V_D \!\equiv\! V$) on resonance.  For $r_{\rm eff} < 2$, one has $K_1^{\rm bare} < 4/(1+r) - 1$, and the system flows to the strong-coupling fixed point at which there is a uniform system and perfect transmission.  
For $r_{\rm eff} > 2$, as the bare coupling $V$ decreases, for instance along the red dashed line, there is a BKT type quantum phase transition at $V\!=\!V^*$. For smaller $V$, resonant tunneling is destroyed, and the flow is toward the decoupled, zero transmission state (blue line of fixed points on the horizontal axis). For $r\!>\!3$ (not shown), the flow is always toward the decoupled state, indicating that resonant tunneling is not possible in this regime. 
}
\label{fig:RG_sym}
\end{figure}

Consider first the special case of symmetric coupling $V_S=V_D \equiv V$ (and still $\epsilon_d \!=\!0$). In this case, $K_2$ is not generated in the RG process, since the RG equation for $K_2$ simplifies to
$ dK_2 / d\ln\tau_c  =  -4\tau_c^2 K_2 V^2 $.
A schematic RG flow diagram is shown in Fig.\,\ref{fig:RG_sym} \cite{Kane92}. There are three regimes: 

(i)~The tunneling $V$ grows under the RG flow and goes to the strong-coupling limit when 
$(1+K_1)(1+r)/4 \!<\!1$ [or equivalently, $K_1<4/(1+r)-1$]. When this is satisfied by $K_1^{\rm bare}$, that is from the beginning of the flow, the physical parameters satisfy
\begin{equation}
r_{\rm eff} \equiv \left[ 1+\Big(\frac{C_G}{C_\Sigma}\Big)^2 \right] r
+4 \Big(\frac{C_G}{C_\Sigma}\Big)^2 r_G < 2 \;.
\label{eq:tostrong}
\end{equation}
For the case $C_G \!\ll\! C_S,C_D$, the criterion for $V$ to grow becomes $r<2$. For the case of only gate coupling ($r\!=\!0$ and $C_\Sigma \!=\! C_G$), $V$ grows if $r_G < 1/2$. 

(ii)~There is the possibility of flow to weak coupling ($V\!=\!0$) when 
$r_{\rm eff}\!>\!2$ and in addition $r\!<\!3$. 
In this case, although large tunneling $V$ flows to strong coupling, as the bare tunneling $V$ decreases a separatrix is crossed, denoted $V^*$, below which $V$ flows to zero. The resonant tunneling is completely destroyed at zero temperature for $V<V^*$; indeed, this flow diagram indicates a Berezinsky-Kosterlitz-Thouless (BKT) type quantum phase transition by tuning the bare tunneling. Note that as $K_1$ scales to $0$, only $r$ appears in the RG equations, suggesting that the gate dissipation becomes unimportant in the very low temperature limit.

(iii)~Finally, the flow of $V$ is always to weak coupling when $r>3$. In this regime, resonant tunneling simply does not occur. 

The ground state at weak coupling [regimes (ii) and (iii)]---for this case of symmetric barriers and exactly on resonance---consists of disconnected source and drain leads plus an uncoupled resonant level \cite{EggertAffleck92}. The conductance is clearly zero. Because the resonant level can be either filled or empty, the ground state is two-fold degenerate. 

As the system flows to strong coupling [regimes (i) and (ii)], the weak-coupling RG is no longer valid, and so we turn to treating a small barrier in order to access the strong-coupling fixed point. It turns out that in this limit as well, our system is equivalent to resonant tunneling in a Luttinger liquid, allowing us to draw on previous results. To show that, it is convenient to use the effective model Eq.\,(\ref{eq:SC_H}) from Sec.\,\ref{sec:mapping_LL} consisting of a double barrier in an effective Luttinger liquid plus an extra density-density interaction, $(d^{\dagger}d-1/2) [\partial_x \widetilde{\phi}_{S}(x=0)+\partial_x \widetilde{\phi}_{D}(x=0)]$. In the strong-coupling limit, the system becomes uniform \cite{EggertAffleck92}, and this operator becomes a density-density interaction in that uniform system, which then has scaling dimension $2$.
Therefore, when the weak-coupling RG flows to strong coupling in regimes (i) and (ii) above, this operator is irrelevant and so can be neglected. 

In the absence of the density-density interaction terms, the effective model Eq.\,(\ref{eq:SC_H}) is exactly the same as that for two barriers in a Luttinger liquid with interaction parameter $g_f$, and so we can immediately use the extensive previous literature \cite{Kane92,*Kane92a,EggertAffleck92,Furusaki93,Sassetti_Napoli_Weiss_95,FendleyPRB95,*FendleyPRL95,MatveevPRB95,FurusakiMatveevPRB95,Furusaki98,FurusakiMatveevPRL02,NazarovGlazman03,polyakov03,komnik03,Meden2005,EnssPRB05,goldstein09,GoldsteinPRL10a}. Note in particular that the parameter $g_c$ and fluctuations involving the gate have disappeared from the problem. The strong-coupling fixed point corresponds to a single, connected, uniform system plus a decoupled fractional degree of freedom \cite{EggertAffleck92,WongAffleck94,AffleckPrivCom}. The transmission is unity for this system. In the special case $r\!=\!1$, the decoupled degree of freedom is a Majorana fermion, and the ground state degeneracy is $\sqrt{2}$, a value familiar from the two-channel Kondo effect with which there is a close tie (see Section \ref{sec:mapping_LL} above).

\subsection{Detuning: Second Quantum Phase Transition}

\begin{figure}[t]
\begin{centering}
\includegraphics[width=2.0in,clip]{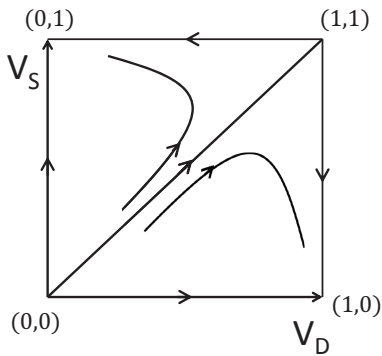}
\par\end{centering}
\caption[Schematic of the RG flow in VsVd plane in regime (i)]
{Schematic representation of the RG flow of the two tunneling amplitudes, $V_{S}$ and $V_{D}$, in regime (i): the level is on resonance and the dissipation is not too 
strong, $r_{\rm eff} \!<\! 2$. The diagonal is the symmetric barrier case: it flows into the strong-coupling quantum critical point at $(1,1)$ which corresponds to a uniform system and so perfect transmission. At point $(1,0)$, the level is fully incorporated into the D lead ($V_D \!=\!1$) while completely disconnected from the S lead ($V_S\!=\!0$); the roles of source and drain are reversed at $(0,1)$. Single barrier scaling is expected along the vertical lines from $(1,1)$ to either $(1,0)$ or $(0,1)$.  
}
\label{fig:RG_asym}
\end{figure}

For the case of asymmetric coupling, $V_S\!\neq\! V_D$, we start with the case $r_{\rm eff}\!<\!2$ [Eq.\,(\ref{eq:tostrong})], namely regime (i) above. For the on-resonance case, the schematic RG flow is shown in Fig.\,\ref{fig:RG_asym}.\cite{Kane92,Furusaki93,EggertAffleck92} First, we consider the weak-coupling RG. As we saw above, along the symmetric line $V_S\!=\!V_D$, the flow is to the strong-coupling fixed point, denoted $(1,1)$, at which one has perfect transmission. For $V_S<V_D$, $V_D$ flows to strong coupling, but $V_S$ flows to zero---point $(1,0)$ in Fig.\,\ref{fig:RG_asym}. This implies complete incorporation of the level into the D lead, but the system is cut in two by the S barrier. For $V_S>V_D$ the two behaviors are interchanged. Thus in the asymmetric coupling case, the zero temperature behavior is to have two disconnected semi-infinite Luttinger liquids, a situation for which the transmission is clearly zero. 

Low temperature properties are determined by the approach to the weakly coupled fixed point $(1,0)$ given by the perturbative RG equations (\ref{eq:RG_VD}). Near this point, the equation for $V_S$ reduces to $d\ln V_S/d\ln\tau_c = -r$.  Thus we see that $G \propto V_S^2 \propto T^{-2r}$ near the weak-coupling fixed point. Note that the gate resistance does not enter this scaling relation; physically, since the level is incorporated into the D lead, charge can flow freely out of the level, and so the gate potential fluctuations have no effect. 

In the vicinity of the strong-coupling fixed point, we note that the double barrier problem can be  mapped onto an effective single barrier problem with effective potential \cite{Kane92}
\begin{equation}
   V_{\rm eff} \cos[\pi (\epsilon_d + 1/2)] \cos(2\sqrt{\pi}\theta)
\end{equation}
where $\theta$ is the plasmon-like displacement field which is dual to $(\widetilde{\phi}_{S} \!+\! \widetilde{\phi}_{D})/2$. The operator here corresponds to $2k_F$ backscattering; we neglect $4k_F$ backscattering (which is irrelevant for $g_f>1/4$) and other higher order process. 

The $2 k_{F}$ reflection vanishes on resonance, $\epsilon_d=0$, for a symmetric double barrier, leading, as mentioned above, to perfect transmission with $G=e^2/h$. (The approach to this value is controlled by operators we have neglected here, 
as discussed in Refs.\,\onlinecite{EggertAffleck92,Huaixiu13_1-G}.) 
A small detuning of $\epsilon_d$ from resonance through an applied gate voltage, $\Delta V_g$, causes a backscattering amplitude that is linear in $\Delta V_g$. Another way to tune away from the unitary resonance is by inducing a slight asymmetry, $V_{S}\neq V_{D}$. In this case, the $2 k_{F}$ backscattering term is proportional to the bare value of $V_{S}-V_{D}$. Thus, the fixed point at $V_S \!=\! V_D$ and $\Delta V_G \!=\! 0$ is unstable in both directions, as observed in the experiment in Refs.\,\onlinecite{Mebrahtu12,Mebrahtu13}. 

Finally, in the off-resonant ($\epsilon_d \!\neq\!0$) weak-coupling case, an extension of the RG equations applies \cite{GoldsteinPRB10}. These show that for asymmetric barriers the behavior off resonance is the same as on resonance, namely flow to a state in which there are two disconnected Luttinger liquid leads. However, in the symmetric barrier case ($V_S\!=\!V_D$ but $\epsilon_d \!\neq\!0$), though the flow is naturally toward weak coupling, the weak-coupling ground state is not the same as in the on-resonant case discussed in Section \ref{sec:SymmOnres} \cite{GoldsteinPRB10,AffleckPrivCom}. 
Here the resonant level is either filled or occupied in the ground state---the ground state is not degenerate. The leading process connecting the two leads is cotunneling via the level; this process is irrelevant, as for tunneling through a single barrier. Thus the system is ultimately cut in two---the source lead and drain lead are disconnected from each other---and the conductance is zero  \cite{GoldsteinPRB10,AffleckPrivCom}. The final state in the off-resonant symmetric case is therefore the same as that in both the resonant and off-resonant asymmetric cases. 

As a function of either asymmetry $V_S \!-\! V_D$ or energy detuning $\Delta V_G$, then, there is a \emph{quantum phase transition} from the fully connected uniform ground state at $(1,1)$ to two disconnected leads. In the experiment of Refs.\,\onlinecite{Mebrahtu12,Mebrahtu13}, this transition and the quantum critical point at $V_S\!=\!V_D$ and $\epsilon_d\!=\!0$ are observed by tuning the couplings and energy level.
Note that at both strong and weak coupling, the effect of barrier asymmetry is similar to that of detuning the resonant level. At strong coupling, both produce backscattering of the same form as scattering from a single (small) barrier. At weak coupling, both cause flow to the case of a single barrier cutting the system. Thus, the scaling is expected to be the same along both directions, a feature seen in the experimental data as well \cite{Mebrahtu12,Mebrahtu13}.
Furthermore, the scaling along the entire vertical line from $(1,1)$ to $(1,0)$ is thought to be given by single barrier scaling \cite{Kane92,EggertAffleck92,AristovWoelfle09,Huaixiu13_1-G}.

\begin{figure}[t]
\begin{centering}
\includegraphics[width=3.3in,clip]{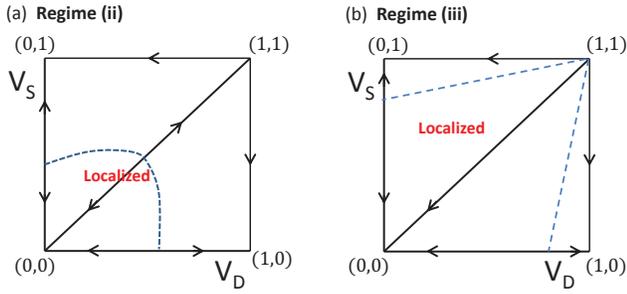}
\par\end{centering}
\caption[Schematic of the RG flow in VsVd plane in regimes (ii) and (iii)]
{Schematic representation of the RG flow of the two tunneling amplitudes, $V_{S}$ and $V_{D}$, when the level is on resonance and the dissipation is strong. (a) Regime (ii), $r_{\rm eff} \!>\! 2$ but $r\!<\!3$. (b) Regime (iii), $r\!>\!3$. The dotted line marks the BKT transition between a localized state in which the level is disconnected from the leads, $(0,0)$, and an extended state in which the level joins seamlessly with either one [(0,1) or (1,0)] or both leads [(1,1))].  
}
\label{fig:RG_asym_strong}
\end{figure}

Turning now to the case of strong dissipation and parameters for which there is 
not flow to strong coupling---namely, in regime (iii) defined above or regime (ii) with $V < V^*$---we see that the asymmetry of the system does not cause a major effect. In the symmetric case, as discussed above in Section \ref{sec:SymmOnres}, there is a BKT transition between the $(0,0)$ disconnected level and the $(1,1)$ uniform system phases. Likewise, in the presence of asymmetry there is a BKT transition between the disconnected level and the (0,1) split system phases. This latter transition has been studied in detail in the context of tunneling to a single lead in the presence of gate dissipation \cite{ButtikerPRL00,FurusakiMatveevPRL02,LeHurLiPRB05,BordaZarandX06,ChungQPT,ChengIngersent09,Goldstein10a}. It corresponds to the classic localized-delocalized transition in the spin boson model \cite{Chakravarty82,LeggettRMP87}. Thus in the $V_S$-$V_D$ plane there is a line along which a BKT transition occurs between a localized and an extended phase: 
Figure \ref{fig:RG_asym_strong} shows schematic RG flows when the level is on resonance for regimes (ii) and (iii). With regard to the flow along the lines (0,0)$\leftrightarrow$(0,1) and (0,0)$\leftrightarrow$(1,0), 
since it is known that for a single lead the delocalized phase appears for any strength of dissipation for sufficiently large $V_D$ [i.e., there is no analog of regime (iii) of the symmetric coupling case] \cite{FurusakiMatveevPRL02}, then the runaway flow from $(1,1)$ to $(1,0)$ always occurs.

\section{Sequential Tunneling} \label{sec:sequential}

We have seen that resonant tunneling is destroyed by dissipation in our system except under very special conditions---the system must have symmetric coupling to the leads and be tuned on resonance. If these conditions are not met, the properties of the system are described by tunneling through a single effective barrier (at low temperature). However, even under the special resonant conditions, resonant tunneling may be destroyed if the dissipation is sufficiently strong---regimes (ii) and (iii) of Section \ref{sec:SymmOnres}. In this case, the low temperature properties of the system are given by sequential tunneling through the localized state. 
In the case of a level embedded in Luttinger liquid, this regime has been analyzed in detail \cite{Furusaki98}. Here, we check that our model of Section \ref{sec:model_dissipative} describes the sequential tunneling regime as well. 

The sequential tunneling regime is treated using rate equations in which the key ingredient is the tunneling rate from the level to each of the leads \cite{IngoldNazarov92,Furusaki98}, in our case $\Gamma_S$ and $\Gamma_D$ for the source and drain leads. These tunneling rates are modified by the coupling to the electromagnetic environment, an effect known as the dynamical Coulomb blockade. We focus on $\Gamma_S$ as an illustration and proceed via two paths, showing that they give the same result: (1) direct calculation from the Hamiltonian Eq.\,(\ref{eq:HT_v2}) and (2) use of standard dynamical Coulomb blockade theory based on the impedance seen from the tunnel junction.

First, coupling to the resistive environment produces a power law suppression in the tunneling rate as a function of temperature; $\Gamma_S = \Gamma_S^0 \, T^{2r'_S}$ defines the exponent $r'_S$. By Fourier transformation, a power law decay in time of the phase correlations as in Eq.\,(\ref{eq:corvarphi}), i.e.\ $t^{-2r_2}$, produces a corresponding dependence on temperature, namely $T^{2r_2}$. Thus, from the Hamiltonian Eq.\,(\ref{eq:HT_v2}) and the correlations of the phase operators $\varphi_2$ and $\varphi_3$ given in Eqs.\,(\ref{eq:R2def})-(\ref{eq:corvarphi}), we find immediately
\begin{eqnarray} \label{eq:rS}
   r'_S & = & r_2/4 + (C_G/C_\Sigma)^2 r_3 /4
   \nonumber \\
   & = & \frac{1}{4}\left[ 1+\Big(\frac{C_G}{C_\Sigma}\Big)^2 \right] r
+ \Big(\frac{C_G}{C_\Sigma}\Big)^2 r_G = \frac{r_{\rm eff}}{4} \;.
\end{eqnarray}

\begin{figure}[t]
\begin{centering}
\includegraphics[width=2.3in,clip]{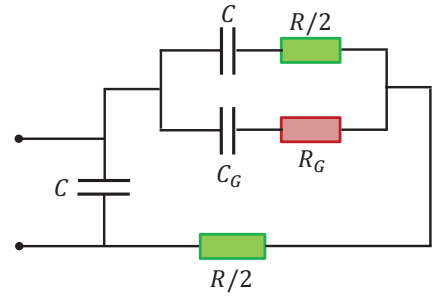}
\par\end{centering}
\caption[Effective circuit for sequential tunneling through S]
{Effective impedance seen by an electron tunneling across the S barrier in the sequential tunneling regime. The real part of this impedance is $r_s' R_Q$ which controls the low temperature scaling of the tunneling rate. 
}
\label{fig:circuit_seq}
\end{figure}

In the second approach, according to dynamical Coulomb blockade theory \cite{IngoldNazarov92}, the temperature dependence of the tunneling rate is controlled by the real part of the low frequency impedance seen between the two sides of the tunneling barrier. The effective circuit is thus shown in Fig.\,\ref{fig:circuit_seq}.\cite{NazarovPrivCom} Indeed, calculating the impedance of this circuit in the low frequency limit yields $Z \approx i/\omega C_\Sigma + r'_S R_Q$ where $r'_S$ is given by the expression above. A simple way to understand the circuit result can be constructed as follows: When an electron tunnels across the source barrier, it causes current in all three branches of the circuit (source, drain, and gate) because of the image charge produced on the three capacitors. The capacitance determines the fraction of the current in each resistor: $C/C_\Sigma$ flows through the drain circuit, $C_G/C_\Sigma$ through the gate, and $C/C_\Sigma$ stays on the source capacitor so that $1-C/C_\Sigma$ flows through the source circuit. Since dissipation is given by the square of the current, we have
\begin{equation}
   r'_S =   \Big(\frac{C}{C_\Sigma}\Big)^2 \frac{r}{2} 
          + \Big(\frac{C_G}{C_\Sigma}\Big)^2 r_G
          + \Big(1-\frac{C}{C_\Sigma}\Big)^2 \frac{r}{2} ,
\end{equation}
which simplifies to the expression in Eq.\,(\ref{eq:rS}). To summarize this section, we see that the approach using the fluctuating modes introduced in Section \ref{sec:model_dissipative} reassuringly reproduces the result of dynamical Coulomb blockade theory. Results for the conductance in the sequential tunneling regime may then be obtained by using rate equations \cite{IngoldNazarov92,Furusaki98}.

\section{Conclusion} \label{sec:conclusion}

In this paper, we investigate the problem of resonant tunneling through a quantum dot in the presence of two dissipative baths, one coming from the resistive source and drain leads and the other from a resistive gate coupled to the energy of the resonant level. We treat a spinless (spin polarized) level relevant for experiment \cite{Mebrahtu12,JezouinPierre13,Mebrahtu13} and consider an electrically source-drain symmetric case, $C_S\!=\!C_D$ and $R_S\!=\!R_D$, though the quantum mechanical tunnel coupling is not necessarily symmetric. The first step is to identify the independent electromagnetic modes which couple to the environment; in our case there are two since the total charge in the dot does not couple. Then, by using bosonization and unitary transformations, we map our problem to several resonant-level Luttinger-liquid-type models. Because of having two distinct dissipative baths, the Luttinger liquid model that results is not of the simplest form (i.e.\,a resonant level embedded in a homogeneous Luttinger liquid) and, in particular, involves two interaction parameters [Eqs.\,(\ref{eq:gfdef}) and (\ref{eq:gcdef})]. Nevertheless, the standard Luttinger liquid tools such as RG based on the Coulomb-gas representation can be used to analyze the new models. We elucidate in what ways our model is similar to the standard Luttinger liquid case and in what ways it differs. 

Two QPT occur in our system, and its different ground states are associated with three RG fixed points that we label (A)-(C). The first QPT occurs for strong dissipation and is of the BKT type. When the resonant level is exactly on resonance with the source and drain leads and is symmetrically coupled to them, this QPT separates (A) a two-fold degenerate state at weak coupling in which the system is cut in two and the level can be either filled or empty [$(0,0)$] from (B) a state in which there is a uniform source-drain system plus a disconnected fractional degree of freedom [$(1,1)$], which for the case $r\!=\!1$ is a Majorana mode thus having a degeneracy of $\sqrt{2}$. 
State (B) incorporates effects similar to those of the two-channel Kondo model, with the two fermionic leads (S and D) acting as different channels. 
When the resonant level is not exactly symmetrically coupled to the leads (but still on resonance), this BKT transition still occurs for sufficiently strong dissipation. It separates state (A) from (C) a state in which the system is simply cut in two with the resonant level incorporated into either the source or drain lead [$(1,0)$ or $(0,1)$].
The existence and nature of this QPT is the same as in the simple resonant level in a Luttinger liquid model. However, crossing or observing this QPT requires strong dissipation. The presence of two dissipative baths in our system eases the criterion needed to observe the BKT transition; the way in which the two baths combine to produce effectively stronger dissipation is given by $r_{\rm eff}$ in Eq.\,(\ref{eq:tostrong}). In addition, the two baths provide a flexibility in parameters that relaxes the constraint $g_f \!=\! g_c$ of the simple Luttinger liquid. 

The second QPT occurs as one tunes away from the special point of symmetric coupling with the level on resonance. Either an asymmetry in the coupling or a detuning of the energy of the resonant level causes the system to flow away from the unusual critical state (B) above to the state (C). The system is cut in two with the resonant level either incorporated into the source or drain lead (asymmetry) or becoming empty or full (level detuning)---these various possibilities are all equivalent. State (C) is not degenerate and is a stable fixed point of the system. We noted that upon approaching both fixed points (B) and (C), the gate dissipation becomes ineffective: the flow is controlled simply by the source-drain dissipation, a situation equivalent to the simple resonant level in a Luttinger liquid model. However, in the full cross-over from (B) to (C), the gate dissipation can be expected to play a significant role. 

The mapping from the dissipative models that we consider to various Luttinger liquid models shows that quantum open systems can be used to emulate 1D interaction effects. This connection has been made explicit in a number of recent  works \cite{Mebrahtu12,JezouinPierre13,Mebrahtu13}. Clearly, this connection can be further developed, leading to ways in which quantum dissipative systems can be used to emulate other more complicated interacting systems. Several extensions of our work come readily to mind: going beyond the electrically symmetric case that we have considered ($C_S\!=\!C_D$ and $R_S\!=\!R_D$), exploring the role of the spin degree of freedom (which has been suppressed here), and studying the scaling near strong-coupling in the case of two baths (what role does the dissipative gate play?). We leave these for future work.

\section*{Acknowledgments}

We thank I. Affleck, S. Florens, and M. Goldstein for valuable discussions. This work was supported by the U.S.\,DOE, Office of Basic Energy Sciences, Division of Materials Sciences and Engineering, under awards \#DE-SC0005237 and \#DE-SC0002765. 
H.U.B. thanks the Fondation Nanosciences de Grenoble for its hospitality while completing this work. 

\vspace*{-0.3in}
\bibliography{QPTbib8,quadruple}

\end{document}